\documentclass[letterpaper]{jpconf}
\usepackage{graphicx}
\begin{document}
\title{Top Quark Physics at the CDF Experiment}

\author{Bernd Stelzer}

\address{Simon Fraser University}

\begin{abstract}
Fermilab's Tevatron accelerator is recently performing at record luminosities that
enables a program systematically addressing the physics of top quarks.
The CDF collaboration has analyzed up to 5 fb$^{-1}$ of proton 
anti-proton collisions from the Tevatron at a center of mass energy of 1.96 TeV. The
large datasets available allow to push top quark measurements to higher and higher precision
and have lead to the recent observation of electroweak single top quark production at the Tevatron.
This article reviews recent results on top quark physics from the
CDF experiment. \end{abstract}

\section{Introduction}
The discovery of the top quark by the CDF and D0 collaborations at Fermilab in 1995 \cite{cdfd0topdisc} 
ended a 20 year quest for the top quark and completed the three-generation structure of the standard model (SM).
The top quark is distinguished by its large mass, close to the scale of electroweak symmetry breaking, 
while all other observed fermions have masses that are a tiny fraction of this energy. 
The origin of this unique property remains mostly a mystery, except that the top quark couples strongly to 
the mechanism of electroweak symmetry breaking (EWSB). It is speculated whether the top quark 
may play a fundamental role in EWSB or is special otherwise.
Studying the top quark in detail may provide hints of new physics.

The large mass of the top quark leads to some interesting features. The top quark decays almost 
exclusively through the single mode $t\rightarrow Wb$. 
The top quark decay is expected to proceed extremely fast, in less than 10$^{-24}$s, which is shorter than the time scale to form hadrons.  
Hence, the top quark provides us with the unique opportunity to study a bare quark.
\section{Production of Top Quarks} 
Top quark production is a rare process at the Tevatron. Roughly only one
in 10 billion inelastic collisions at the Tevatron features top quarks while the rate for generic QCD jet production is many orders of 
\begin{figure}[!h]
\begin{center}
  \includegraphics[width=.95\textwidth]{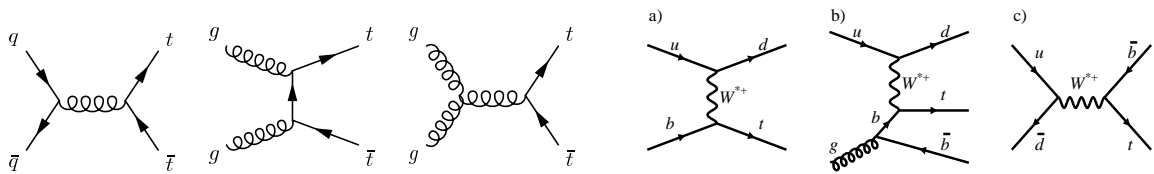}
   \caption{(Left) Representative Feynman diagrams of top quark pair production.
    (Right) $t$-channel (a, b) and $s$-channel (c) single top quark production.}
 \label{fig:cdftprod}
\end{center}
\end{figure}
magnitude larger. Within the SM, top quarks are produced in pairs via the strong interaction 
and singly via the electroweak interaction as shown in Figure \ref{fig:cdftprod}.

\subsection{Measurement of Top Quark Pair Production} 
The theoretical production cross section of top quark pairs has been calculated at Next to Leading Order (NLO) and is
$\sigma_{t\bar{t}}=7.4^{+0.5}_{-0.7} $~pb at the Tevatron, assuming a top quark mass of 172.5 GeV/c$^2$ \cite{ttbarTheory}.
The presence of two top quarks gives rise to unique event signatures classified by the decay mode of the $W$ bosons. 
The most important (or golden) channel is the lepton + jets channel where one $W$ boson decays leptonically into a charged lepton ($e,\mu$) and a neutrino, while 
the second $W$ boson decays into quarks that form hadronic jets. This channel features a large branching ratio of about 34\% with manageable backgrounds, 
and allows for the full reconstruction of the event kinematics. The channel with the highest signal purity is the dilepton channel. 
Both $W$ bosons decay to a lepton ($e,\mu$) and a neutrino. The branching ratio for dilepton events is about 6\% and backgrounds
are small. More challenging channels, due to large backgrounds, are the all-hadronic channel with a branching ratio of about 46\% and 
events with hadronic tau decays that contribute with a branching ratio of about 14\% in tau + jets and tau + lepton final states.
\begin{figure}[h]
\begin{center}
\includegraphics[width=.50\textwidth]{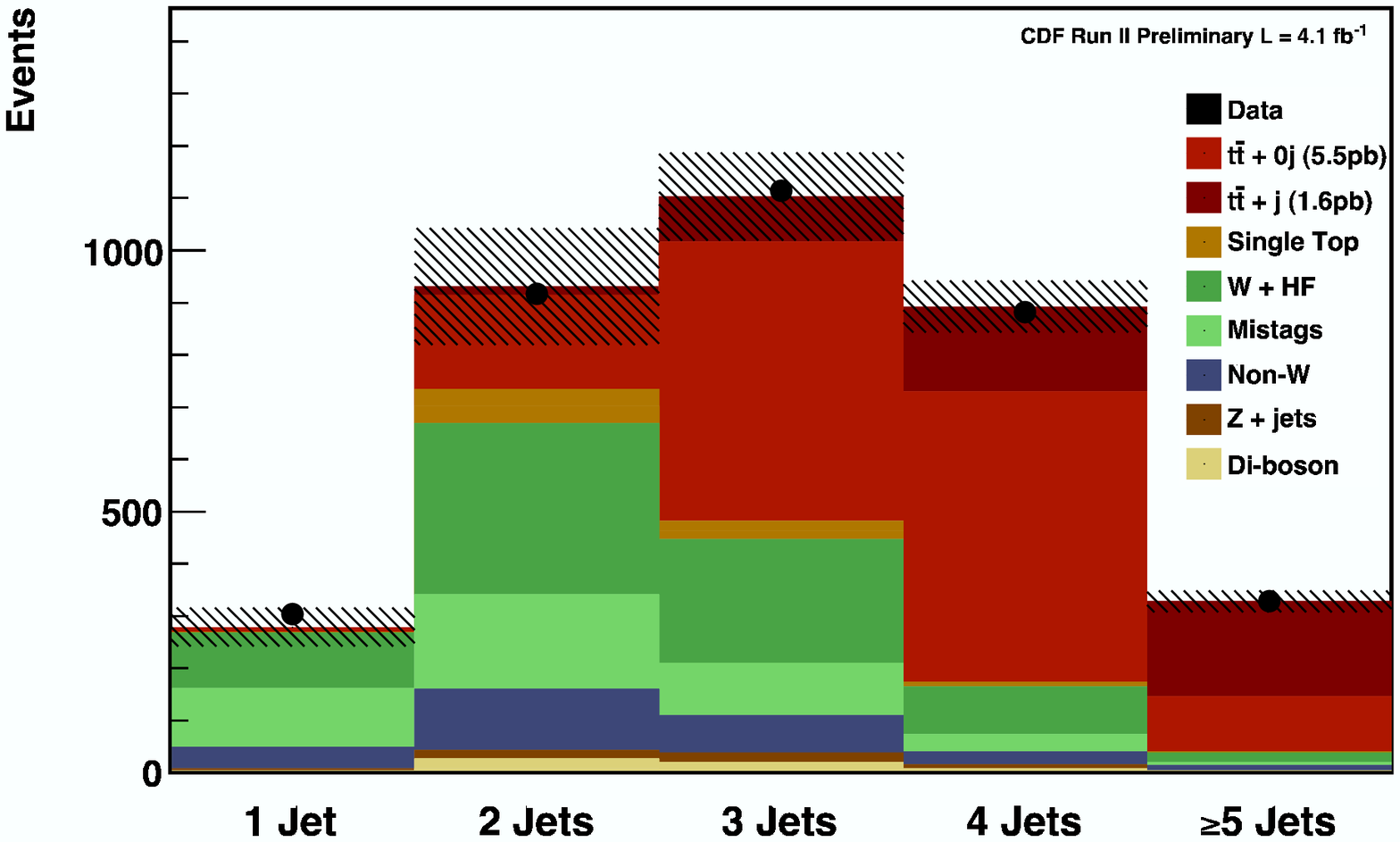}
\hspace{0.3cm}
\includegraphics[width=.40\textwidth]{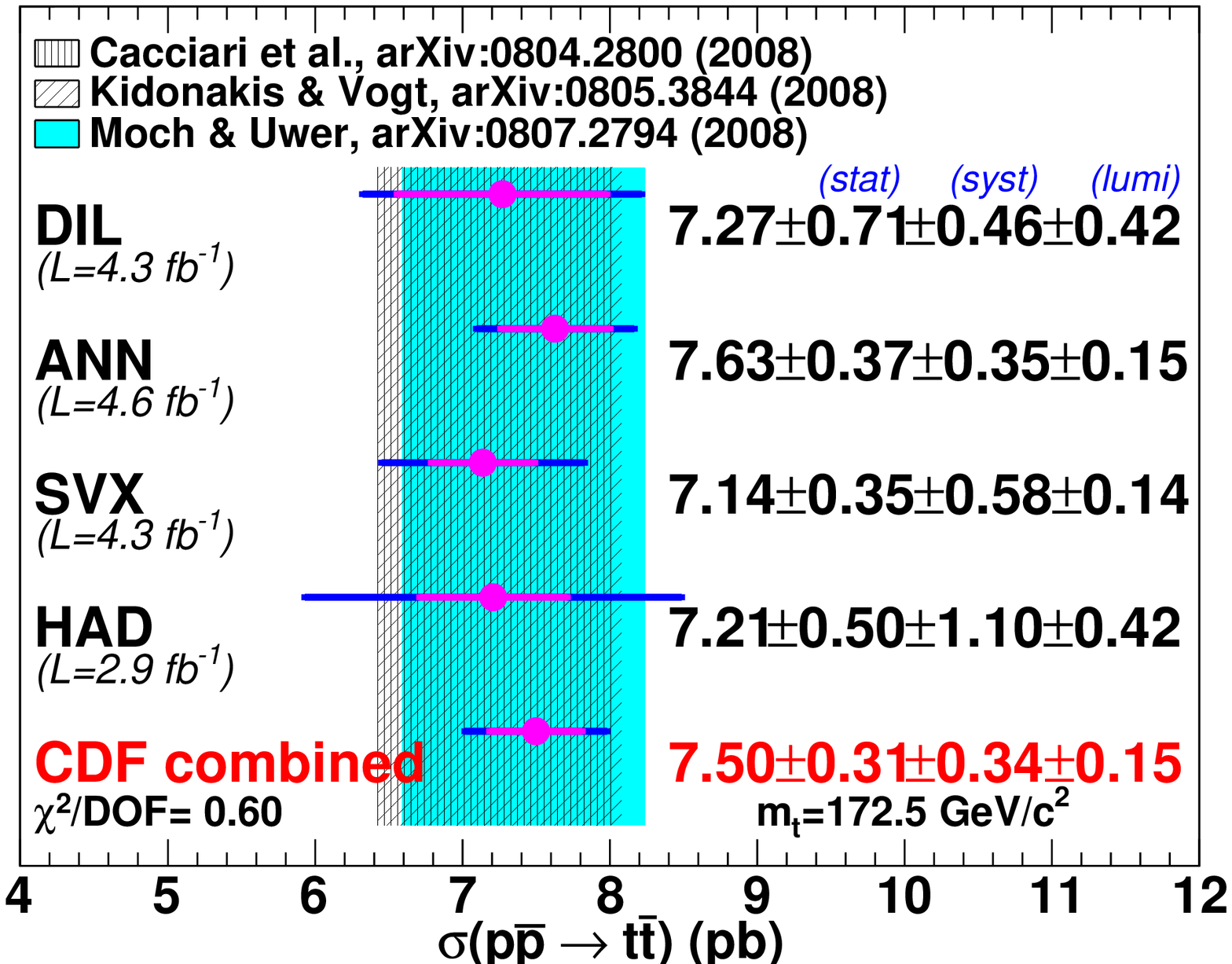}
\caption{(Left) Measurement of top pair production cross section in the lepton + jets channel. (Right) Summary of recent measurements, assuming a top quark mass of 172.5 GeV/c$^2$.}
 \label{fig:cdfttxsecs}
\end{center}
\end{figure}

Measuring the $t\bar{t}$ production cross section is important because a significant difference between measurements
and the theoretical prediction would indicate the presence of new physics. 
CDF has recently updated four measurements using the largest available datasets, close to 5 fb$^{-1}$.
One measurement was performed in the dilepton channel (DIL), one measurement in the all-hadronic channel (HAD) and two measurements 
in the lepton+jets channel. The two measurements in the lepton+jets channel employ different strategies for
signal extraction. One uses an artificial neural network based on seven kinematic and topological observables (ANN).
The other analysis uses displaced secondary vertex $b$-tagging to purify the sample (SVX).
The latter analysis is essentially a counting experiment in each jet multiplicity bin of the $W$ + Jets distribution
as shown in Figure \ref{fig:cdfttxsecs}.
Both measurements are systematically limited. The dominating uncertainty on luminosity (6\%) has recently been 
reduced by measuring the ratio of the top quark pair cross section to the cross section of $Z$ boson production and
normalizing to the theoretical $Z$ boson cross section. This makes both analyses nearly free of the luminosity uncertainty.

The new CDF measurements are summarized on the right of Figure \ref{fig:cdfttxsecs}.
The results indicate consistency among different final states and consistency with the theoretical prediction.
The combination of all measurements yields $\sigma_{t\bar{t}} = 7.50 \pm 0.31 (stat) \pm 0.34 (syst) \pm 0.15 (Z_{theory})$~pb
assuming a top quark mass of 172.5 GeV/c$^2$.
\subsection{Observation of Single Top Quark Production} 
Top quarks are also expected to be produced singly in electroweak interactions
through a $s$-channel or $t$-channel exchange of a virtual $W$ boson, as
shown on the right of Figure~\ref{fig:cdftprod}. The combined NLO cross section for both channels
has been calculated to be $\sigma_{st}=2.86\pm 0.36$~pb, assuming a top quark mass of 175 GeV/c$^2$~\cite{stxs}.
The weak coupling strength of the top quark is not very well constrained,
except that $|V_{tb}|^2\gg |V_{td}|^2+|V_{ts}|^2$ \cite{topR}. Requiring that the
$3\times 3$ CKM matrix is unitary implies that $|V_{tb}|\simeq 1$~\cite{pdg}.
With a matrix of higher rank, though, $|V_{tb}|$ could be small without 
measurably changing the $t\rightarrow Wb$ branching ratio. 
Production of single top quarks provides a direct measurement of $|V_{tb}|$.

The smaller production cross section of single top quarks and the presence of only one 
top quark in the final state make the separation of the signal events from large 
backgrounds very challenging. Using improved lepton, jet and $b$-quark 
identification algorithms allows the selection of a candidate sample with a signal to background ratio of 
about $\sim$~1/20.
\begin{figure}[h!]
\begin{center}
  \includegraphics[width=.80\textwidth]{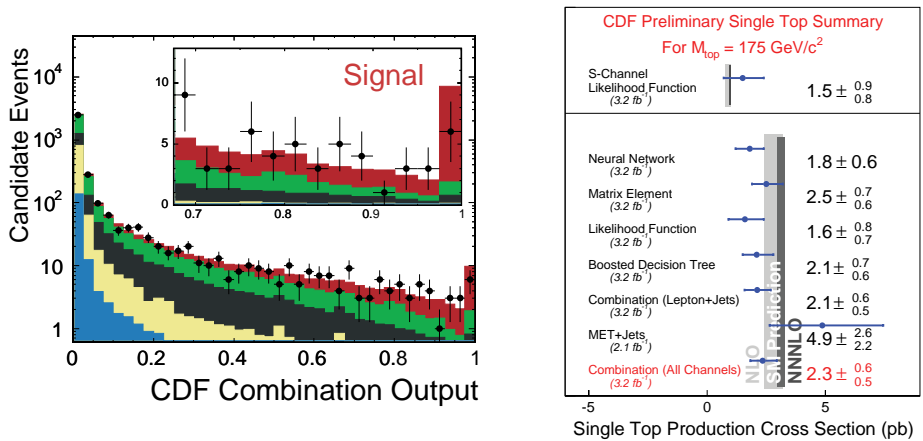}
   \caption{(Left) Combined discriminant output distribution of the single top quark observation at CDF.
 (Right) Summary of single top cross section results using different techniques.}
 \label{fig:cdfST}
\end{center}
\end{figure}
This poor signal to background ratio renders a simple counting experiment impossible
and demands for sophisticated signal extraction techniques;
a matrix element method, boosted decision trees, neural networks and a likelihood
function approach have been developed at CDF. The large top quark mass and angular correlations in top quark decay
makes this process very well suited for multivariate analyses.
Careful validation of these techniques in data control samples have been performed.
The CDF collaboration announced (jointly with the D0 experiment) the observation of single top
quark production at the 5.0 standard deviations level in March 2009 \cite{stobservation}. Figure \ref{fig:cdfST} on the left shows the
discriminant distribution, obtained from a combination of several analyses summarized on the right of Figure \ref{fig:cdfST}. 
The combined cross section measured by CDF is $\sigma_{st}=2.3^{+0.6}_{-0.5}$~pb consistent with the SM prediction.
The cross section result translates into $|V_{tb}|=0.91\pm0.11 (exp) \pm 0.07 (theory)$, the most precise direct determination of $V_{tb}$ (by a single
experiment) to date.
\section{Top Quark Properties} 
In the past year, the CDF collaboration has produced more than 30 new top quark property measurements and searches for new
physics in the top quark sample. A few will be highlighted in this section.
\subsection{Top Quark Mass}
The mass of the top quark is a fundamental parameter of the standard model. When combined with the 
$W$~boson mass measurement, it places constraints on the mass of the unobserved Higgs boson. 
\begin{figure}[h!]
\begin{center}
  \includegraphics[width=.8\textwidth]{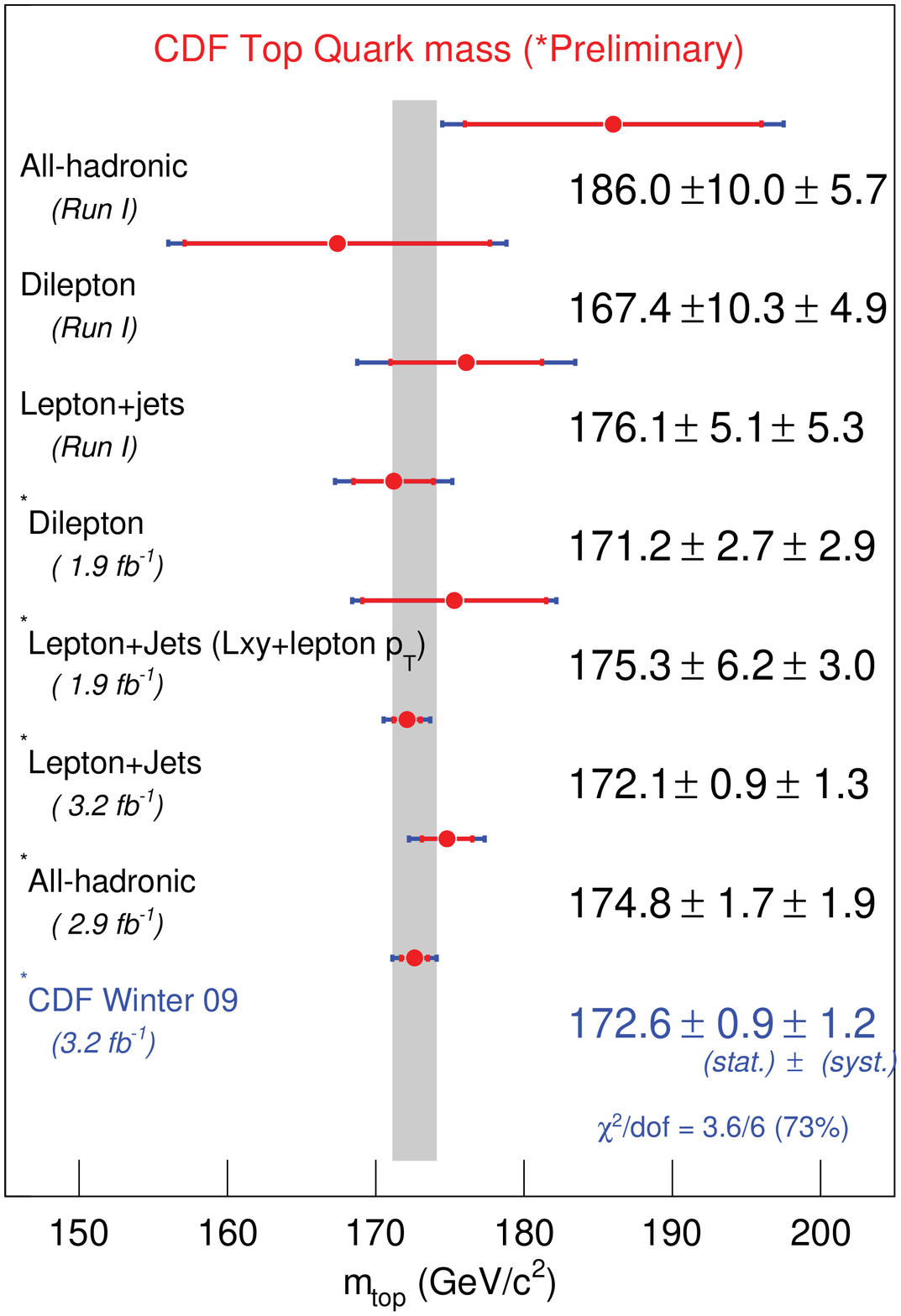}
   \caption{(Right) Summary of top quark mass measurements at CDF. (Left, top) Measurement of top quark mass in lepton + jets channel. (Left,
   bottom) Distribution of forward backward asymmetry in top pair events $-Q_l\cdot y_{had}$. }
 \label{fig:cdfmass}
\end{center}
\end{figure}
CDF has measured the top quark mass in all top pair decay channels using different analysis
techniques. Experimental challenges arise from determination of the jet energy scale, the combinatorial 
ambiguity to assign measured jets to the parton of the hard interaction as well as complications due to
QCD radiation.  
For improved sensitivity, the jet energy scale is constrained, {\em in-situ}, through the hadronic decay of 
the $W$ boson present in the top quark decay. A summary of all top quark mass measurements at CDF is 
shown in Figure~\ref{fig:cdfmass}. The most precise measurements are obtained in the lepton + jets channel.
A top quark mass template fit is shown on the left of Figure~\ref{fig:cdfmass}.
It is worth noting that results among different channels and measurement techniques are consistent.
The CDF combined result is $m_t = 172.6 \pm 0.9(stat) \pm 1.2(syst)$ GeV/c$^2$ and has a precision of less than 0.7\%. 
There is an ongoing effort at the Tevatron to improve and unfold sources of systematic uncertainty in collaboration with 
the theory community~\cite{tevTopMass}.

\subsection{Top Quark Spin and Lifetime}
One of the most remarkable properties of the top quark is its extremely short lifetime.
Top quarks are not polarized in its pair production mode, though there are correlations between the top and
anti-top quark spin since the production is mediated through a spin-1 gluon.
The spin correlation coefficient can be extracted from data, in the dilepton channel, 
by performing a fit to the double differential distribution $\cos\theta_1 \cos\theta_2$ of the
positive and negative lepton direction of flight in the top quark decay.
Using 2.8 fb$^{-1}$ of data, CDF measures a value of $\kappa = 0.32^{+0.55}_{-0.78}$ using the off-diagonal spin
quantization axis, consistent with the standard model prediction of $\kappa = 0.78$ \cite{ofdbasis}. A consistent result is
obtained in an analysis using lepton + jets events measured in the helicity basis. CDF has also set direct limits
on the top quark width using a modification of the top quark mass analysis and obtained 
$\Gamma_{top}<$~7.5 GeV/c$^2$ at 95\% C.L.

\subsection{Forward Backward Asymmetry} 
Quantum chromodynamics at leading order predicts that 
the top quark production angle is symmetric with respect to the beam direction. 
At NLO, QCD predicts a small charge asymmetry, $A_{fb} = 0.050 \pm 0.015$ \cite{AfbTheory}, 
due to interference of initial and final-state radiation
diagrams and interference of box diagrams with the Born level process. 
In the CP invariant Tevatron frame, the charge
asymmetry is equivalent to a forward-backward asymmetry of the 
produced top quarks. CDF measures a forward-backward asymmetry 
in the proton anti-proton lab frame of $A_{fb}^{det} = 0.193 \pm 0.065 (stat) \pm 0.024 (syst)$.
The CDF result indicates some tension with the standard model prediction and confirms
an earlier measurement \cite{cdfAfb}. New physics could give rise to an increased asymmetry 
through interference of new particles at high energy not yet directly observed. More data 
will be required to differentiate a statistical fluctuation from new physics.

\subsection{Search for Resonance Top Quark Production}
Several models of new physics predict resonance production of a new exotic
particle that decays predominantly into top quark pairs \cite{maltoni}. 
Resonant top quark production ($X\rightarrow t\bar{t}$) is possible through 
massive new particles predicted in extended gauge theories, Kaluza Klein states of the gluon or Z boson, 
axigluons and topcolor. Independent of the exact model, a narrow width resonance should be visible in the
$t\bar{t}$ invariant mass distribution. 
In two analyses using data corresponding to 2.8 fb$^{-1}$ of data, the CDF collaboration 
found no evidence for such a resonance in the lepton + jets channel and all hadronic channel. 
In the absence of any signal CDF places upper limits on $\sigma_X \times BR(X\rightarrow t\bar{t})$.
If interpreted in the context of a topcolor-assisted technicolor model \cite{topcolor} these limits can be used to derive mass 
limits on a narrow lepto-phobic $Z^{\prime}$ with $M_{Z^{\prime}} >$ 805 GeV/c$^2$ at the 95\% C.L., 
assuming $\Gamma(Z^{\prime})$ = 0.012 $M_{Z^{\prime}}$.

\section{Conclusions}
The discovery of the top quark opened up a rich field in particle physics.
A broad top physics program is underway at CDF \cite{topTevatron}.
So far, the data seems consistent with the standard model.
While the precision of the top quark pair production cross section
has surpassed the theoretical prediction, several top quark properties measurements
are still statistically limited and will benefit from the larger datasets
expected at the Tevatron and the Large Hadron Collider in the near future.
The observation of single top quark production provides a new sample
to study top quark properties and probe physics beyond the standard model \cite{ttait}.

\end{document}